\documentclass[useAMS,usenatbib]{mn2e}
\usepackage{amssymb,graphicx,natbib,longtable,hyperref,breakurl}
\bibpunct{(}{)}{;}{a}{}{,}

\usepackage[utf8]{inputenc}
\usepackage[english]{babel}

\def \ls61 {LS~I~+61$^{\circ}$303 }

\title[\ls61 observed with $Swift$/BAT]{Temporal features of \ls61 in hard X-rays
from the $Swift$/BAT survey data}

\author[A. D'A\`i et al.]{A. D'A\`i$^{1}$, \thanks{E-mail:antonino.dai@ifc.inaf.it}
G. Cusumano$^{1}$,
V. La Parola$^{1}$, 
A. Segreto$^{1}$,
T. Mineo$^{1}$\\
$^{1}$ INAF - Istituto di Astrofisica Spaziale e Fisica Cosmica, Via U.\ La Malfa 153, I-90146 Palermo, Italy.
}

\begin{document}

\pagerange{\pageref{firstpage}--\pageref{lastpage}} \pubyear{}
\maketitle

\label{firstpage}
\begin{abstract}
We study the long-term spectral and  timing behaviour of \ls61 in hard
X-rays (15--150  keV) using $\sim$\,10  years of survey data  from the
$Swift$  Burst  Alert  Telescope  (BAT)  monitor.   We  focus  on  the
detection of long periodicities known to  be present in this source in
multiple wavelengths. We clearly detect three periods: the shorter one
at 26.48 days is compatible with the orbital period of the system; the
second, longer, periodicity  at 26.93 days, is detected  for the first
time in X-rays and its value  is consistent with an analogous temporal
feature recently detected in the  radio and in the gamma-ray waveband,
and we  associate it with a  modulation caused by a  precessing jet in
this  system.  Finally,  we  find   also  evidence  of  the  long-term
periodicity at  $\sim$\,1667 d,  that results  compatible with  a beat
frequency of the two close, and shorter, periodicities. We discuss our
results in  the context  of the multi-band  behaviour of  the physical
processes of this source.
\end{abstract}

\begin{keywords}
X-rays: binaries -- X-rays: individual: LS~I~+61$^{\circ}$303.
\noindent
Facility: {\it Swift}
\end{keywords}

\section{Introduction} \label{intro}
\ls61  is an  accreting  binary system  well-known  for exhibiting  an
exceptional   broadband   spectrum   from  radio   to   TeV   energies
\citep{albert08}.  The system consists of a main sequence star of type
B0 Ve \citep{hutchings81}, with  an estimated mass 10--15 M$_{\odot}$,
and a compact object (it is still debated if a black-hole or a neutron
star) orbiting with  a period $P_{\rm orb}$\,=\,26.4960\,$\pm$\,0.0028
d  in a  highly  eccentric orbit  ($e$\,=\,0.537\,$\pm$\,0.034), at  a
distance from us of 2 kpc \citep{aragona09}.

The  system  is characterized  by  different  long periodicities:  the
shortest one is associated with the above-mentioned orbital period and
it  is detected  in all  bands of  the electromagnetic  spectrum, from
radio, where it was first noticed \citep{taylor82} up to $\gamma$-rays
\citep{abdo09}.   The longest  periodicity,  also first  noted in  the
radio  band \citep{gregory99},  is clearly  super-orbital at  a period
$\sim$\,4.6 years (P$_{\rm so}$), and  because of the larger time-span
required to  detect it, it has  been only recently detected  at higher
frequencies  \citep[see][and reference  therein]{li14}.  In  the X-ray
band, the modulation appears phase shifted of $\sim$\,0.2 with respect
to the radio one \citep{li12}.  In between there is a periodicity very
close     to    the     known     orbital     period    at     $P_{\rm
  2}$\,=\,26.99\,$\pm$\,0.08 d,  that has been more  recently observed
only  in  the radio  and  in  the $\gamma$-ray  bands  
\citep{massi13, jaron14} and,  finally, a  periodicity that  appears as  an averaged
value   between   $P_{\rm   2}$   and   $P_{\rm   orb}$   at   $P_{\rm
  av}$\,=\,26.704\,$\pm$\,0.004, that has  been exclusively attributed
to the radio outburst recurrence time \citep{ray97, jaron13}.

According to \citet{massi14} $P_{\rm 2}$  is caused by a precession of
a  conical  jet, as  also  revealed  by  the  periodic change  of  the
associated extended radio structure \citep{massi12}, while the $P_{\rm
  so}$ is  the result  of the  beat between  the two  shorter periods,
$P_{\rm  orb}$  and $P_{\rm  2}$  \citep{massi13}:  namely, its  value
(within the  statistical uncertainties)  results compatible  with this
hypothesis    ($P_{\rm    so}$\,=\,($P_{\rm    orb}^{-1}$\,--\,$P_{\rm
  2}^{-1}$)$^{-1}$).   Other   authors  suggest  instead   a  possible
connection   with  the   time-scales  of   the  Be   stellar  activity
\citep{ackermann13},  or the  precession  of the  Be companion  star's
decretion disk \citep{lipunov94}.

In this paper,  we exploit the continuous hard X-ray  coverage of this
source made in  the last $\sim$\,10 years by  the $Swift$/BAT monitor,
to  assess the  presence  and the  spectral  characteristics of  these
periodicities in the hard X-ray band.

\section{Data reduction and analysis}
\begin{figure}
\begin{center}
\includegraphics[width=\columnwidth]{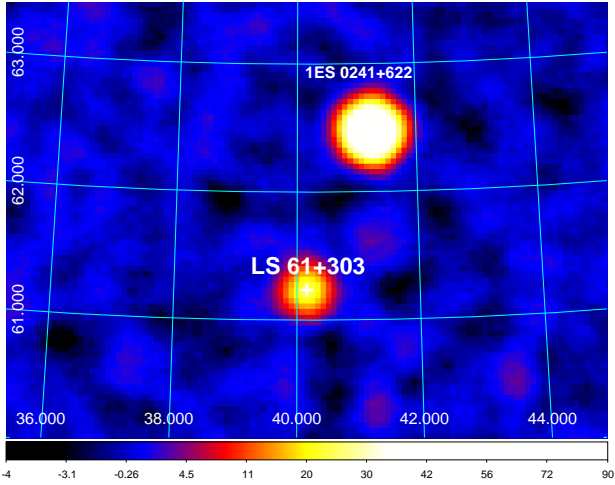} 
\caption[BAT sky map]{The 15--150 keV Swift/BAT image of the \ls61
  sky region.  The brightest  nearby object,  the quasar  0241+622, at
  $\sim$\,80\arcmin results clearly detached.}
\label{map}
\end{center}
\end{figure}

We  retrieved the  survey data  for \ls61  collected with  $Swift$/BAT
between 2004  December 09 (MJD  53348) and  2015 March 10  (MJD 56827)
from                 the                HEASARC                 public
archive\footnote{\url{http://heasarc.gsfc.nasa.gov/docs/archive.html}}
and processed them using a  software package dedicated to the analysis
of  the data  from coded  mask telescopes  \citep{segreto10}. The  BAT
light curve consists  of 55,312 entries and the source  was in the BAT
field-of-view $\sim$\,15 times  a day, for $\sim$\,412 s  each time on
average.  The  source is  detected with  high significance  ($\sim$\,25
standard  deviations) in  the  15--150 keV  BAT range.  Fig.~\ref{map}
shows the significance map of the BAT sky region around LS~I~+61$^{\circ}$303.

Data  analysis was  performed using  the HEASARC/FTOOLS  v. 6.16,  the
RStudio  software  \citep{rstudio}  and the  specific  Lomb-Scargle  R
package \citep{ruf99} We report errors  at one sigma confidence level,
unless stated otherwise.

\subsection{Temporal analysis: detection of the $P_{\rm so}$}
We first  studied the long-term light  curve of the BAT  data, looking
for the $P_{\rm  so}$ presence.  Because the  data cover approximately
slightly more than  two complete orbits, we tried to  directly fit the
light  curve using  as  a best-fitting  model the  sum  of a  constant
emission and a sine function, assuming the shape of the periodicity is
sinusoidal.   Data  were  re-binned  using  a  time-bin  of  80  days,
corresponding to about three complete  orbital periods. The values for
the constant rate,  the sine amplitude, the period and  the phase were
initially all left as free parameters.

\begin{figure}
\includegraphics[height=\columnwidth, angle=-90]{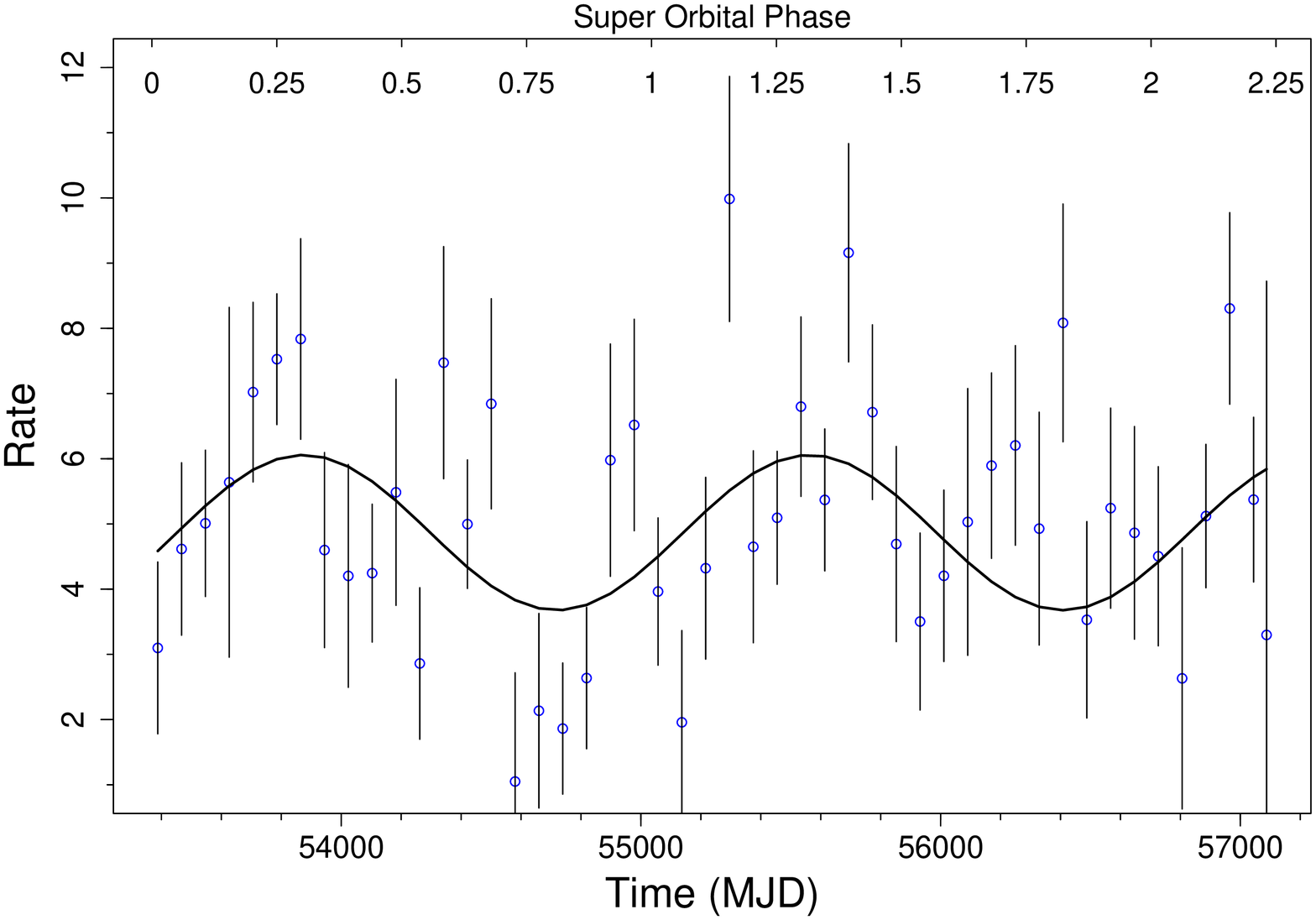}
\includegraphics[height=\columnwidth, angle=-90]{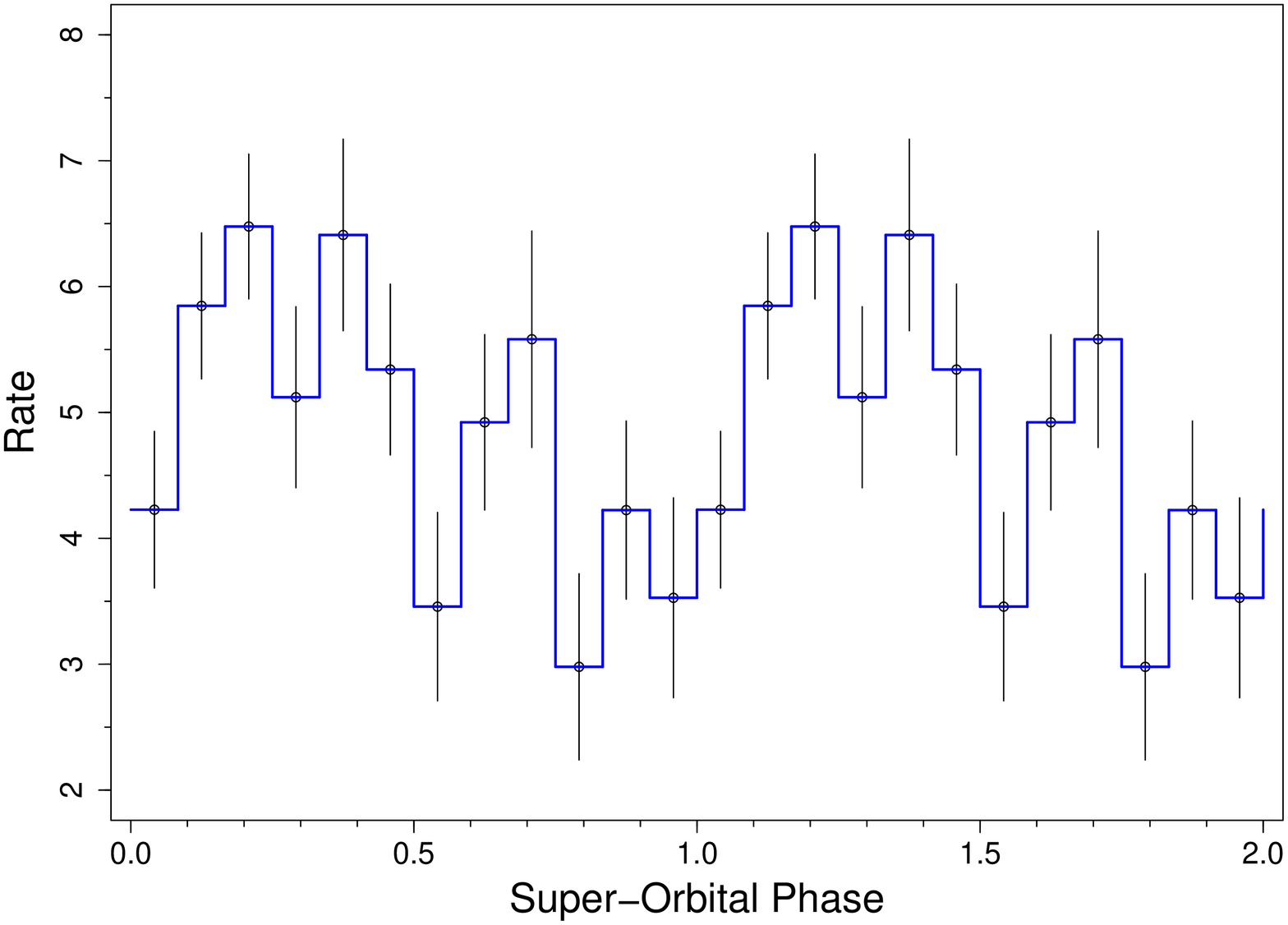}
\caption{ {\it Upper panel}:  $Swift$/BAT light curve (15--150 keV
  range)  with  over-imposed  the  best-fitting  long-term  modulation
  $P_{\rm  so}$. Time  is in  MJD (bottom  axis) and  in super-orbital
  phase (upper axis).   {\it Lower panel}: folded  BAT light curve
  (12  bins) at  period $P_{\rm  so}$\,=\,1667 d,  with time  zero MJD
  43,366.275. Two consecutive periods are  shown for clarity.  The BAT
  rate  in  both panels  is  in  units  of 10$^{-5}$  counts  s$^{-1}$
  pixel$^{-1}$.
\label{lc_bat}}
\end{figure}

We  found an  averaged flux  of (4.9\,$\pm$\,0.2)\,$\times$\,10$^{-5}$
counts     s$^{-1}$     pixel$^{-1}$,      a     semi-amplitude     of
(1.2\,$\pm$\,0.3)\,$\times$\,10$^{-5}$ counts  s$^{-1}$ pixel$^{-1}$ a
super-orbital period P$_{\rm  so}$\,=\,1689\,$\pm$\,112 d. We show in the upper 
and lower panel of Fig.\ref{lc_bat} the BAT light curve with super-imposed the 
best-fitting model and the folded profile at $P_{\rm so}$, respectively.  The F-test
that compares this model with  the null-hypothesis of no modulation in
the  data  gives  $\sim$\,1\%  probability  that  the  improvement  is
obtained  only by  chance. The  sinusoid peaks  at the  super-orbital
phase $\sim$\,0.2,  compatibly   with  the   phase  shift   ($\Delta
\phi$\,=\,0.17\,$\pm$\,0.02) observed in soft X-rays and in hard X-ray
with $INTEGRAL$/ISGRI data \citep{li12, li14}.

We then  extracted two,  statistically similar,  energy-selected light
curves in the 15--35 keV  (source significance $\sim$ 20 $\sigma$) and
in the  35--150 keV (significance  $\sim$ 16 $\sigma$) bands  to check
the  amplitude dependence  on energy.   To this  aim, we  obtained the
values  for  best-fitting  function  composed  of  a  constant  and  a
sinusoidal  component  as  previously  described. We  found  that  the
modulation   is  statistically   detected  in   the  softer   band  at
$>$\,3\,$\sigma$  ($P_{\rm  so}$\,=\,1715\,$\pm$\,140  and  amplitude
fraction     $\sim$\,0.3)     while    only     marginal     detection
($\sim$\,2\,$\sigma$) is obtained for the harder band.

\subsection{Temporal analysis: Lomb-Scargle periodograms}

\begin{figure}
\includegraphics[height=\columnwidth, angle=-90]{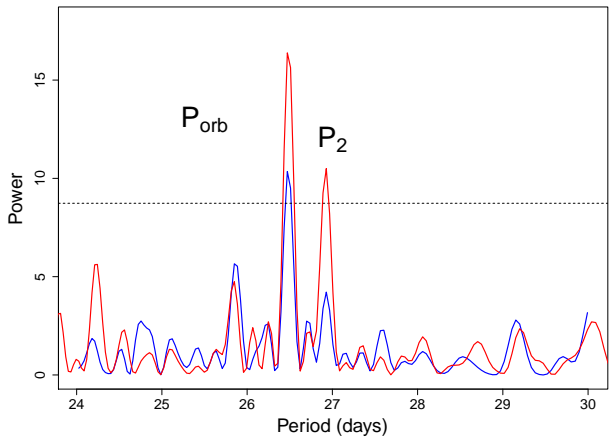}
\caption{LSP  of the  filtered BAT  data-set  in the  22--32 d  period
  region  of interest.   False  alarm  probability (horizontal  dotted
  line) is set at 0.01. Blue (red) line is the LSP for the BAT dataset
  with 23\% (35\%) of noisiest data removed. \label{lsp01}}
\end{figure}

We searched for any periodicity in the 22--32 d period range using the
Lomb-Scargle periodogram (LSP)  technique \citep{lomb76}.  We consider
that the  error on the detected  periods is the half-width  of the bin
period,  that  is  for  the  BAT  data-set  and  periods  of  interest
$\sim$\,0.10 d.  A  preliminary search using the whole  dataset in the
15-150 keV band did not result in any significant detection.  In fact,
LSP method is insensitive to  the statistical error associated to each
measure, while the BAT survey data  are characterized by a wide spread
of  non-Gaussian statistical  errors  that depend  on several  factors
(mainly  the  reduction of  the  coded  fraction  when the  source  is
observed at large  off-axis angles). In this case,  a filtering method
may help a weak feature to  emerge over the noise.  We therefore begun
to gradually remove data with the largest associated rate uncertainty.
We noted that  by filtering out from the original  dataset the 23\% of
the   data   with   the   largest  uncertainty,   a   periodicity   at
26.47\,$\pm$\,0.10  d  starts  to  be  significantly  detected,  while
removing the 35\% of the noisiest data, resulted in the detection of a
second  period  of  slightly  higher value  at  26.93\,$\pm$\,0.10  d,
compatible with the $P_{\rm 2}$ period that had been reported in radio
and in the gamma band (see Fig.~\ref{lsp01}).

We repeated  the same procedure  for the two  energy-filtered datasets
(15--35  keV and  35--150  keV  bands). We  noted  again  that it  was
necessary  to   remove  part  of   the  noisiest  entries   to  obtain
statistically significant  detections. As for the  entire energy range
the orbital  period is the  first feature  to emerge, followed  by the
$P_{2\rm }$  detection.  In  particular, for  the softest  band $both$
periods are detected when 36\% of  the data are removed, while for the
hardest energy band this happens after 24\% of the data are removed.

\begin{figure}
\includegraphics[height=\columnwidth, angle=-90]{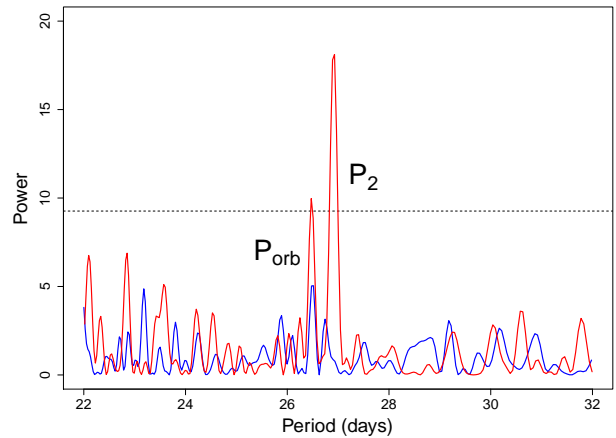}
\caption{LSP of  two phase-selected datasets.  Blue (red) line  is the
  LSP for the  orbital phase interval 0--0.5  (0.5--1.0).  False alarm
  probability (horizontal dotted line) set at 0.01. \label{lsp02}}
\end{figure}

Following \citet{jaron14},  we then passed  to study the power  of the
two signals for different orbital  phase intervals (where the phase is
calculated assuming  as time  of reference  $T_{\rm 0}$\,=\,43,366.275
MJD and $P_{\rm orb}$\,=\,26.496 d), using a moving window of constant
phase width of 0.5, with no selection on energy, and filtering out the
noisiest  data  when needed.  We  found  that  the intensity  of  both
features strongly depends on the orbital phase selection: this is more
clearly demonstrated by the two  phase intervals around the periastron
(0\,$<$  $\Phi$ $<$\,0.5)  and  apoastron  (0.5\,$<$ $\Phi$  $<$\,1.0)
passages.  We show  in Fig.~\ref{lsp02}  the corresponding  LSPs: data
belonging to  phase $\Phi<$\,0.5 (blue  line) do not show  evidence of
any  periodicity (irrespective  of other  additional filters  based on
energy and/or  rate error), whereas  data belonging to  phase interval
$\Phi>$\,0.5  (red line)  clearly show  both $P_{\rm  2}$ and  $P_{\rm
  orb}$ periodicities  (when 35\% of  noisiest data are  removed).  We
note that the power associated  to the $P_{\rm 2}$ periodicity becomes
sensibly stronger with respect to  $P_{\rm orb}$, in analogy with what
observed also in the GeV band \citep{jaron14}.

Finally, we studied the temporal evolution of the signals according to
the super-orbital phase.  To this  aim, we selected 5 phase intervals,
choosing the  boundaries so to keep  the same number of  data for each
interval (i.e. 0, 0.15, 0.31, 0.54,  0.79, 1).  We found that both the
$P_{\rm 2}$  and the $P_{\rm orb}$  periods could be well  detected in
the LSP only for the  super-orbital phase interval 0.15--0.31, whereas
marginal   significant   detection   is   obtained   for   the   other
phase-intervals.

\section{Orbital modulation of the spectral shape}
\begin{figure}
\begin{tabular}{c}
\includegraphics[height=\columnwidth, angle=-90]{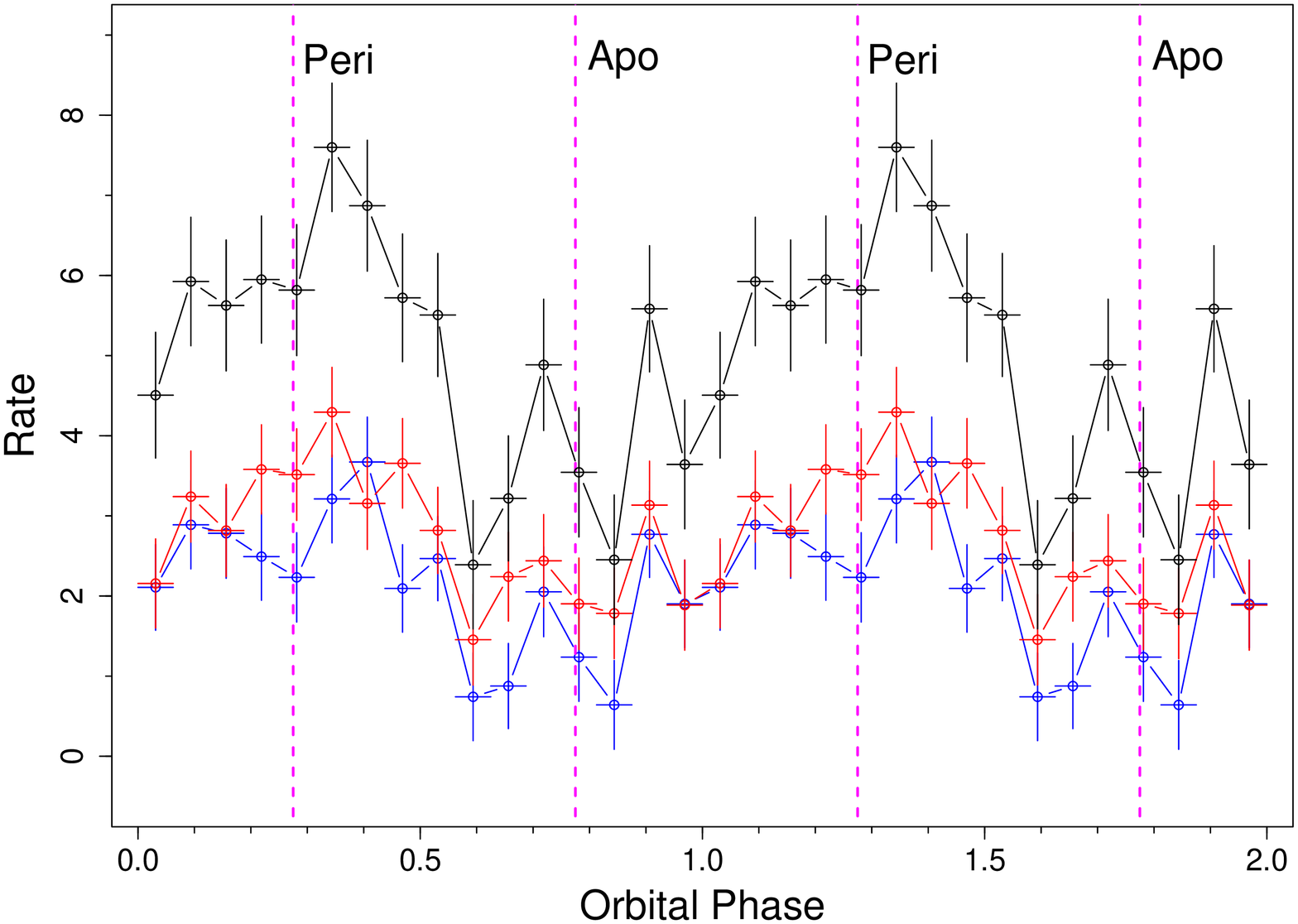} \\
\includegraphics[height=\columnwidth, angle=-90]{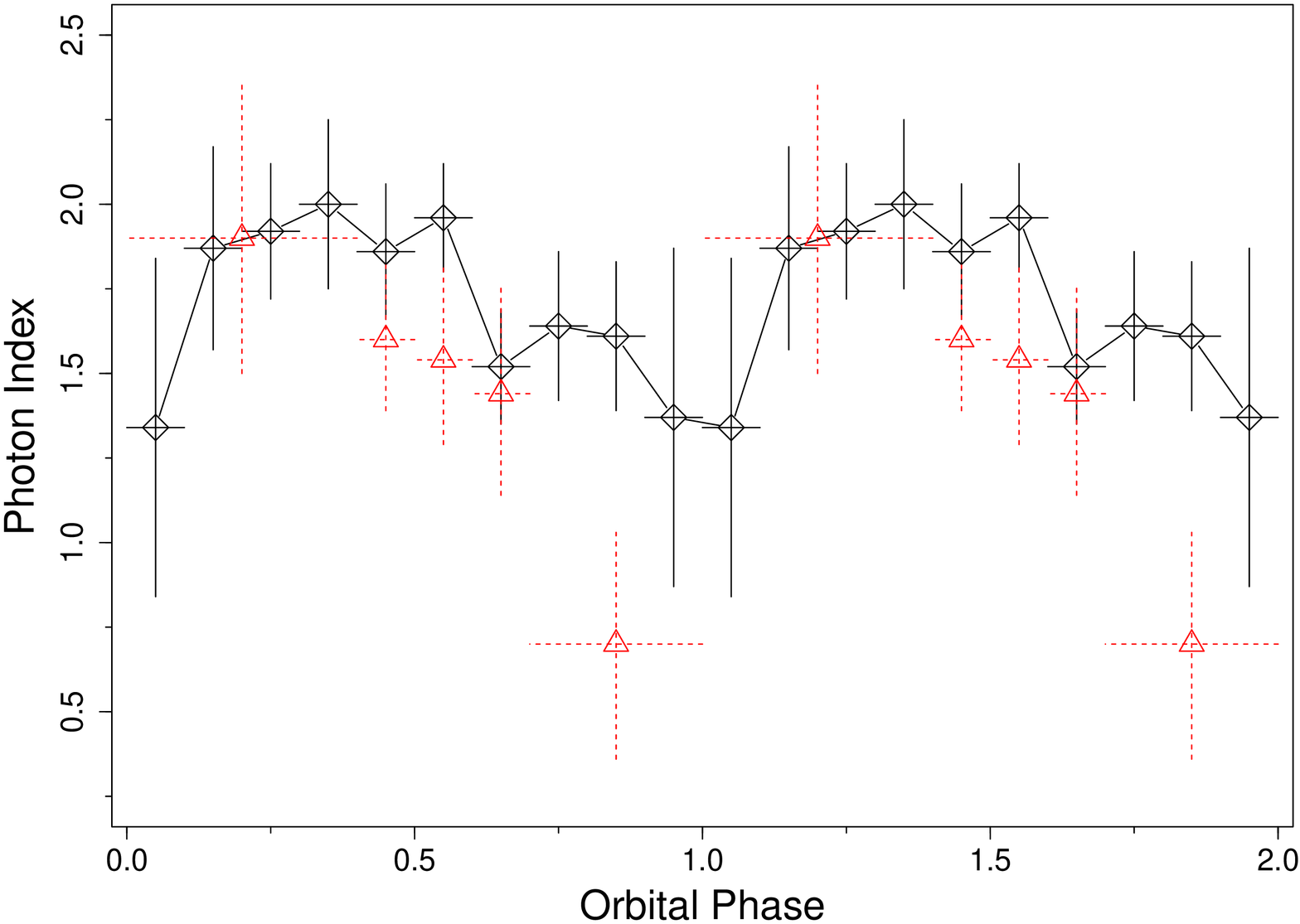} 
\end{tabular}
\caption{{\it Upper panel}: folded BAT light curve (15--110 keV range)
  at  P$_{\rm   orb}$\,=\,26.496  d.   Time  zero   of  reference  MJD
  43366.275.  Red,  blue, and  black curves are  data selected  in the
  15--35  keV,  35--150  keV  and 15--150  keV  ranges,  respectively.
  Magenta  dotted lines  indicate the  periastron and  apoastron phase
  passages \citep{aragona09}.  The  BAT rate is in  units of 10$^{-5}$
  counts s$^{-1}$  pixel$^{-1}$.  {\it Lower panel}:  the photon-index
  of the power-law that best fits the data in the 15--150 keV range as
  a function  of the orbital phase.   We also show for  comparison the
  values   obtained  with   $INTEGRAL$/ISGRI  data   (red triangles)
  according to \citet{li14}. Two periods shown for clarity.
\label{fig:folded}}
\end{figure}

We studied the spectral shape of the hard X-ray emission as a function
of the orbital,  $P_{\rm orb}$, and jet precession  $P_{\rm 2}$ phase.
In  the upper  panel of  Fig.~\ref{fig:folded}  we show  the BAT  data
folded at  $P_{\rm orb}$ for  three energy bands: 15--35  keV, 35--150
keV,  and 15--150  keV.  The emission  peaks
at  phase $\sim$\,0.3, while the phase of minimum emission appears more structured around the 
apoastron passage. The observed maximum flux ratio is  $\sim$\,6.  The folded profile is similar
in the two energy-selected  bands, although  it is to be noted that the
soft emission is  enhanced over the hardest band in  the first half of
the  orbital cycle.  This is  most  easily observed  through a  direct
spectral  fit of  the  phase-selected spectra.  We  choose 10  equally
spaced phase selected spectra and fitted them using a simple power-law
model.  We  show  in  the lower  panel  of  Fig.~\ref{fig:folded}  the
dependence of the photon-index as a function of the orbital phase.  We
observe a clear trend as a function of the phase that gives an account
of the overall modulation of the energy-selected folded profiles.

\begin{figure}
\includegraphics[height=\columnwidth, angle=-90]{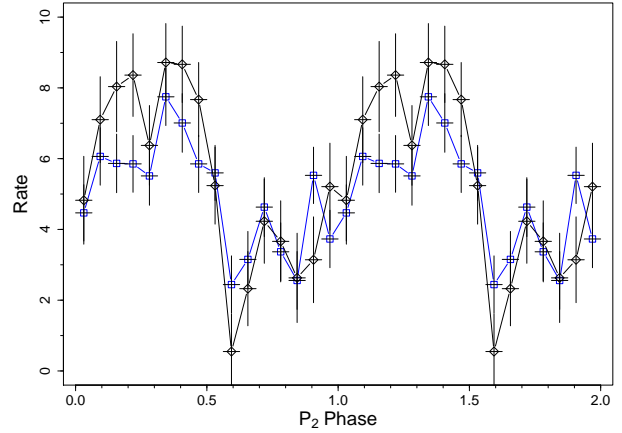} 
\caption{Folded   BAT    light   curve   (15--110   keV    range)   at
  P$_{2}$\,=\,26.93 d.  Time zero  of reference MJD 43366.275. Profile
  in  blue is  phase  averaged (see  LSP  of Fig.~\ref{lsp01}),  while
  profile in  black is from data  selected in the $P_{\rm  orb}$ phase
  ($\Phi>$\,0.5, see LSP  of Fig.~\ref{lsp02}).   The BAT rate  is in
  units          of         10$^{-5}$          counts         s$^{-1}$
  pixel$^{-1}$. \label{fig:folded2}}
\end{figure}

We show in  Fig.~\ref{fig:folded2} two folded profiles  at $P_{\rm 2}$
using the  same epoch of reference  of the folded $P_{\rm  orb}$ and a
period of 26.93 d, that is the  our best value according to the LSP of
Fig.~\ref{lsp02}. The  profile in  blue is  averaged over  all orbital
phases,  whereas the  profile in  black  is obtained  when the  signal
becomes enhanced in the orbital phase interval 0--0.5.

\section{Discussion}
We  examined  the $Swift$/BAT  light  curve  of  \ls61 to  assess  the
presence   and  the   spectral   characteristics   of  its   long-term
periodicities.  \citet{li14}  reported the  presence of  the long-term
$P_{\rm  so}$   modulation  in  hard  X-rays   using  $INTEGRAL$/ISGRI
data. The ISGRI energy band (17--60 keV) almost entirely overlaps with
the  BAT band,  and the  temporal window  used in  that analysis  (MJD
52579--56500)  covers $\sim$\,84\%  of the  BAT window,  so it  can be
considered  an excellent  benchmark  for our  results  (but the  ISGRI
exposure is only  10\% of the BAT one). We  confirm that the long-term
$P_{\rm so}$ is well detected (at $\sim$\,4\,$\sigma$) in hard X-rays.
The BAT $P_{\rm so}$ folded  profile (Fig.~\ref{lc_bat}) is similar to
that  obtained  from  ISGRI,  although the  larger  and  more  uniform
distribution of the exposures at all phases allows us to derive a more
detailed profile. We confirm a  higher X-ray emission during the first
half of the super-orbital cycle.

We  assessed  the  presence  of  the $P_{\rm  orb}$  and  $P_{\rm  2}$
periodicities in  the 15--150 keV  band that  are also visible  in two
different energy  bands (15--35 keV  and 35--150 keV).  In particular,
the detection of $P_{\rm 2}$ \citep[already revealed only in radio and
  in the $\gamma$ band,][]{massi15}) is reported for the first time in
X-rays.  We found this feature more  prominent in the data selected in
the  orbital  phase  0.5--1.0,  and this  behaviour  matches  the  GeV
analogous feature  \citep{jaron14, massi15}; we consider  this finding
as a  strong indication for  a common  origin for the  apastron energy
emission  mechanism,  from  radio,   X-rays  to  the  GeV  wavelength.
\citet{massi14}  proposed  a  physical  scenario able  to  relate  the
presence of these periodicities across such different energy bands. In
this picture,  the long-term  period is  a beat  period caused  by the
combination  of  a precessing  jet  (at  a  period $P_{\rm  2}$)  that
receives  a modulated  fraction of  plasma with  a slightly  different
period  (the  $P_{\rm  orb}$).    The  overall  emission  (synchrotron
emission emitted by  relativistic electrons in the  magnetized jet) is
the highest  when the  jet forms  the minimum angle  with our  line of
sight  and its  emission  becomes Doppler  boosted.  We estimated  the
expected beat period ($\nu_{\rm beat}  = \nu_{\rm 2} - \nu_{\rm orb}$)
of these two  periodicities to be 1560\,$\pm$\,480  d, consistent with
our measure of the long-term modulation $P_{\rm so}$, although we note
the rather large  uncertainties on $P_{\rm 2}$ and  $P_{\rm orb}$.  We
found that the  power of the two signals depends  on the super-orbital
phase, and  it is maximum  in both  cases for the  super-orbital phase
0.15--0.31, that  corresponds to the  peak of the $P_{\rm  so}$ folded
profile (lower panel of Fig.\ref{lc_bat}).

We also studied the spectral shape in hard X-rays as a function of the
orbital phase ($P_{\rm orb}$) for the whole BAT time-span. The folded,
time-averaged over the whole BAT observing window, X-ray profile peaks
close to the periastron passage, and it shows 
two dips, before and after the apostron, that hint for a secondary peak at this phase (Fig.~\ref{fig:folded}, upper panel). 
Although  the ISGRI  folded  profile  and  the values  of  the
photon-index appear to  be quite similar \citep[see Tab.\,1 in][]{li14},  
BAT data allow a  more detailed account  of the photon-index variation  along the
orbit  and suggest for the  second half  of  the orbital  cycle
(Fig.~\ref{fig:folded},  lower  panel) softer  indices.

Finally, we also reported for the  first time the folded X-ray profile
at the presumed jet periodicity, that shows a significant amplitude in
the flux  emission, comparable  to that shown  in the  orbital period,
when phase-selected around the apoastron passage.

\section{Acknowledgements}
\small 
This work was supported in Italy by ASI contract I/004/11/1.
\bibliographystyle{mn2e}
\bibliography{refs}

\begin{thebibliography}{}

\bibitem[\protect\citeauthoryear{{Abdo}, {Ackermann} \& {Ajello}}{{Abdo}
  et~al.}{2009}]{abdo09}
{Abdo} A.~A.,  {Ackermann} M.,    {Ajello} M.,  2009, \apjl, 701, L123

\bibitem[\protect\citeauthoryear{{Ackermann}, {Ajello} \& {Ballet}}{{Ackermann}
  et~al.}{2013}]{ackermann13}
{Ackermann} M.,  {Ajello} M.,    {Ballet} J.,  2013, \apjl, 773, L35

\bibitem[\protect\citeauthoryear{{Albert}, {Aliu}, {Anderhub} \&
  {Antoranz}}{{Albert} et~al.}{2008}]{albert08}
{Albert} J.,  {Aliu} E.,  {Anderhub} H.,    {Antoranz} P.,  2008, \apj, 684,
  1351

\bibitem[\protect\citeauthoryear{{Aragona}, {McSwain}, {Grundstrom}, {Marsh},
  {Roettenbacher}, {Hessler}, {Boyajian} \& {Ray}}{{Aragona}
  et~al.}{2009}]{aragona09}
{Aragona} C.,  {McSwain} M.~V.,  {Grundstrom} E.~D.,  {Marsh} A.~N.,
  {Roettenbacher} R.~M.,  {Hessler} K.~M.,  {Boyajian} T.~S.,    {Ray} P.~S.,
  2009, \apj, 698, 514

\bibitem[\protect\citeauthoryear{{Gregory}, {Peracaula} \& {Taylor}}{{Gregory}
  et~al.}{1999}]{gregory99}
{Gregory} P.~C.,  {Peracaula} M.,    {Taylor} A.~R.,  1999, \apj, 520, 376

\bibitem[\protect\citeauthoryear{{Hutchings} \& {Crampton}}{{Hutchings} \&
  {Crampton}}{1981}]{hutchings81}
{Hutchings} J.~B.,  {Crampton} D.,  1981, \pasp, 93, 486

\bibitem[\protect\citeauthoryear{{Jaron} \& {Massi}}{{Jaron} \&
  {Massi}}{2013}]{jaron13}
{Jaron} F.,  {Massi} M.,  2013, \aap, 559, A129

\bibitem[\protect\citeauthoryear{{Jaron} \& {Massi}}{{Jaron} \&
  {Massi}}{2014}]{jaron14}
{Jaron} F.,  {Massi} M.,  2014, \aap, 572, A105

\bibitem[\protect\citeauthoryear{{Li}, {Torres} \& {Zhang}}{{Li}
  et~al.}{2014}]{li14}
{Li} J.,  {Torres} D.~F.,    {Zhang} S.,  2014, \apjl, 785, L19

\bibitem[\protect\citeauthoryear{{Li}, {Torres}, {Zhang}, {Hadasch}, {Rea},
  {Caliandro}, {Chen} \& {Wang}}{{Li} et~al.}{2012}]{li12}
{Li} J.,  {Torres} D.~F.,  {Zhang} S.,  {Hadasch} D.,  {Rea} N.,  {Caliandro}
  G.~A.,  {Chen} Y.,    {Wang} J.,  2012, \apjl, 744, L13

\bibitem[\protect\citeauthoryear{{Lipunov} \& {Nazin}}{{Lipunov} \&
  {Nazin}}{1994}]{lipunov94}
{Lipunov} V.~M.,  {Nazin} S.~N.,  1994, \aap, 289, 822

\bibitem[\protect\citeauthoryear{{Lomb}}{{Lomb}}{1976}]{lomb76}
{Lomb} N.~R.,  1976, \apss, 39, 447

\bibitem[\protect\citeauthoryear{{Massi} \& {Jaron}}{{Massi} \&
  {Jaron}}{2013}]{massi13}
{Massi} M.,  {Jaron} F.,  2013, \aap, 554, A105

\bibitem[\protect\citeauthoryear{{Massi}, {Jaron} \& {Hovatta}}{{Massi}
  et~al.}{2015}]{massi15}
{Massi} M.,  {Jaron} F.,    {Hovatta} T.,  2015, \aap, 575, L9

\bibitem[\protect\citeauthoryear{{Massi}, {Ros} \& {Zimmermann}}{{Massi}
  et~al.}{2012}]{massi12}
{Massi} M.,  {Ros} E.,    {Zimmermann} L.,  2012, \aap, 540, A142

\bibitem[\protect\citeauthoryear{{Massi} \& {Torricelli-Ciamponi}}{{Massi} \&
  {Torricelli-Ciamponi}}{2014}]{massi14}
{Massi} M.,  {Torricelli-Ciamponi} G.,  2014, \aap, 564, A23

\bibitem[\protect\citeauthoryear{{Ray}, {Foster}, {Waltman}, {Tavani} \&
  {Ghigo}}{{Ray} et~al.}{1997}]{ray97}
{Ray} P.~S.,  {Foster} R.~S.,  {Waltman} E.~B.,  {Tavani} M.,    {Ghigo} F.~D.,
   1997, \apj, 491, 381

\bibitem[\protect\citeauthoryear{{RStudio Team}}{{RStudio
  Team}}{2015}]{rstudio}
{RStudio Team} 2015, RStudio: Integrated Development Environment for R.
RStudio, Inc., Boston, MA

\bibitem[\protect\citeauthoryear{Ruf}{Ruf}{1999}]{ruf99}
Ruf T.,  1999, Biological Rhythm Research, 30, 178

\bibitem[\protect\citeauthoryear{{Segreto}, {Cusumano}, {Ferrigno}, {La
  Parola}, {Mangano}, {Mineo} \& {Romano}}{{Segreto} et~al.}{2010}]{segreto10}
{Segreto} A.,  {Cusumano} G.,  {Ferrigno} C.,  {La Parola} V.,  {Mangano} V.,
  {Mineo} T.,    {Romano} P.,  2010, \aap, 510, A47

\bibitem[\protect\citeauthoryear{{Taylor} \& {Gregory}}{{Taylor} \&
  {Gregory}}{1982}]{taylor82}
{Taylor} A.~R.,  {Gregory} P.~C.,  1982, \apj, 255, 210

\end{thebibliography}
\end{document}